\documentclass[12pt,preprint]{aastex}
\usepackage{apjfonts}
\slugcomment{Accepted for publication in ApJL}
\shorttitle{Early Optical Polarization of Forward Shock Afterglow of GRB 091208B}
\shortauthors{Uehara et al.}
\begin{document}
\title{
GRB 091208B: First Detection of the Optical Polarization
in Early Forward Shock Emission of a Gamma-Ray Burst Afterglow
}
\author{T.  \textsc{Uehara}\altaffilmark{1},
          K. Toma\altaffilmark{2},
          K.  S. Kawabata\altaffilmark{3}, 
          S. Chiyonobu\altaffilmark{1},
          Y. Fukazawa\altaffilmark{1}, 
          Y. Ikejiri\altaffilmark{1}, 
          T. Inoue\altaffilmark{4},
          R. Itoh\altaffilmark{1}, 
          T. Komatsu\altaffilmark{1},
          H. Miyamoto\altaffilmark{1},
          T. Mizuno\altaffilmark{3},
          O. Nagae\altaffilmark{1}, 
          H. Nakaya\altaffilmark{5},
          T. Ohsugi\altaffilmark{3},
          K. Sakimoto\altaffilmark{1},
          M. Sasada\altaffilmark{1}, 
          H. Tanaka\altaffilmark{1}, 
          M. Uemura\altaffilmark{3}, 
          M. Yamanaka\altaffilmark{3,1},
          T. Yamashita\altaffilmark{4}, 
          R. Yamazaki\altaffilmark{4}, and
          M. Yoshida\altaffilmark{3}          
                    }

\altaffiltext{1}{Department of Physical Science, Hiroshima University, Kagamiyama 1-3-1, Higashi-Hiroshima 739-8526, Japan; uehara@hep01.hepl.hiroshima-u.ac.jp} 
\altaffiltext{2}{Department of Earth and Space Science, Osaka University, Toyonaka 560-0043, Japan}
\altaffiltext{3}{Hiroshima Astrophysical Science Center, Hiroshima University, Higashi-Hiroshima, Hiroshima 739-8526, Japan} 
\altaffiltext{4}{Department of Physics and Mathematics, Aoyama Gakuin University, Fuchinobe, Chuou-ku, Sagamihara 252-5258, Japan}
\altaffiltext{5}{National Astronomical Observatory of Japan, Osawa, Mitaka, Tokyo 181-8588, Japan}
\begin{abstract}
We report that the optical polarization in the afterglow of GRB~091208B is measured
at $t=149 - 706\;$s after the burst trigger, and the polarization degree is
$P = 10.4\% \pm 2.5\%$.
The optical light curve at this time shows a power-law decay with index 
$-0.75 \pm 0.02$, which is interpreted as the forward shock synchrotron emission, 
and thus this is the first detection of the early-time optical polarization in the forward 
shock (rather than that in the reverse shock reported by \citep{Steele09}).
This detection disfavors the afterglow model in which 
the magnetic fields in the emission region are random on the plasma skin depth scales,
such as amplified by the plasma instabilities, e.g., Weibel instability. 
We suggest that the fields are amplified by the magnetohydrodynamic instabilities, 
which would be tested by future observations of the temporal
changes of the polarization degrees and angles for other bursts.
\end{abstract}


\section{Introduction}
\label{sec:intro}
Many of the gamma-ray burst (GRB) afterglows can be explained 
as the synchrotron emission from the shock produced by the 
interaction of the ejecta with the circumburst medium
\citep[][for reviews]{zhang04,piran04}, and their spectra and light curves
have helped us understand the total energy scale, the outflow structure, 
and the circumburst medium profile as well as the microphysical conditions 
of the relativistic collisionless shock for 
each afterglow \citep[e.g.,][]{panaitescu02}.

Major problems on the physics of the relativistic collisionless shocks 
involve how to accelerate particles into the population with the power-law 
energy distribution $dn/d\gamma \propto \gamma^{-p}$, where $\gamma$ is
the electron Lorentz factor, and how to amplify the magnetic field from the typical
strength in the interstellar medium $\sim 1\;\mu$G
to the strength required to produce bright synchrotron radiation 
$\sim 1\;$G for GRB afterglows. 
Polarimetric observations of synchrotron radiation 
from the relativistic shocks can reveal the magnetic field structure.
The field structure is essential for constraining the mechanisms of field 
amplification and particle acceleration (see also \citealt{Toma08}).
The extensive polarimetric observations of the late-time optical afterglows 
($t \sim 1\;$day) have been performed \citep{covino04},\footnote{
Here and hereafter we define $t$ as the time after the burst trigger.} 
although the field structure is still in debate \citep[for a review, see][]{lazzati06}. 
Early polarimetric observation is crucial.

Polarimetric observation generally needs large amount of 
photons and therefore needs a larger telescope than typical 
imaging, because the required accuracy is mostly of $\leq 1$\% order.
To obtain polarimetric data of early afterglow of GRBs,
we should have a polarimeter of which the field of view 
is larger than the initial position error of GRBs ($\sim 3'$) 
as well as a quickly moving telescope.
Since the early emission of GRBs changes rapidly, the 
polarimeter should also have a function to obtain all Stokes parameters
for linear polarization, $I, Q,$ and $U$ within a short timescale.

Recently, optical polarimetry has been performed at the early phase.
\cite{Mundell07} reported an upper limit on the polarization degree of 
GRB 060218 as $P < 8\%$ at $t \gtrsim $ 203~s, corresponding to the 
onset phase of the forward shock. \cite{Steele09} detected
an optical polarization of $P = 10~\% \pm 1~\%$ at $t \gtrsim $ 161~s 
{\it in the reverse shock emission} of GRB~090102.
Both GRBs were observed by RINGO attached to the 2.0~m robotic
Liverpool Telescope, which uses a rotating polarizer. 

In this Letter, we report the first detection of the early forward shock 
optical polarization for GRB~091208B with the Kanata 1.5~m telescope
at Higashi-Hiroshima Observatory. 
The optical emission is explained as 
the forward shock synchrotron emission, and the polarization degree is 
$P = 10.4\% \pm 2.5\%$ averaged over $t = 149$--$706\;$s.
The polarization of the forward shock emission provides us with the information 
on the structure of the magnetic field amplified from the weak circumburst 
magnetic field, rather than possible magnetic fields advected from the central 
engine which may be probed with the polarization from the reverse shock 
\citep{Steele09}. 
It also might be interesting to compare our data for the optical afterglow with 
the recent claim of detection of $\gamma$-ray 
polarization $P_{\gamma} = 27\% \pm 11\%$ 
in the prompt emission of GRB~100826A \citep{yonetoku11}.

The prompt emission of GRB~091208B has the duration $T_{90} = 14.9 \pm 3.7\;$s,
and the fluence in the $15$--$150\;$keV band $f_{\gamma} = (3.3 \pm 0.2) \times
10^{-6}\;{\rm erg}\;{\rm cm}^{-2}$. The redshift is determined as $z \simeq 1.063$,
so that the isotropic $\gamma$-ray energy is estimated
as $E_{\gamma,{\rm iso}} \simeq 1 \times 10^{52}\;$erg \citep{pagani10}.

\section{Optical Data: Observations and Reduction}
\label{obs:091208b}

\begin{figure}
  \begin{center}
    \begin{tabular}{c}
      \resizebox{85mm}{!}{\includegraphics{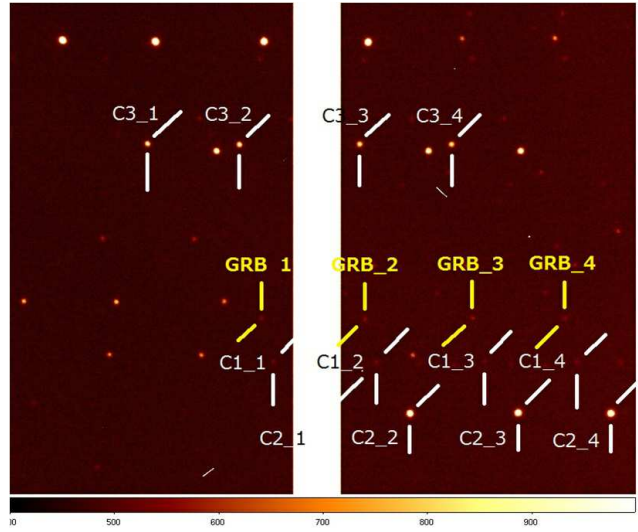}} \\
    \end{tabular}
    \caption{A sample image obtained by HOWPol in one-shot polarimetry mode for GRB~091208B.
 Each object produces four images by linearly polarized rays at 0$^{\circ}$, 
 90$^{\circ}$, 45$^{\circ}$ and 135$^{\circ}$ position angles, respectively, on the projected sky.
 C1--C3 are comparison stars for the magnitude reference. The vertical gap
 around the center ($\sim 40''$ width) is due to the mechanical gap between the two CCDs.}
  \label{fig:img091208b}
    \end{center}
\end{figure}

We performed optical imaging polarimetry of GRB~091208B with HOWPol
(Hiroshima One-shot Wide-field Polarimeter; \citealt{Kawabata08})
attached to the Nasmyth focus on 1.5-m Kanata telescope at
Higashi-Hiroshima Observatory, Japan.
Since HOWPol uses a wedged double Wollaston prism \citep{Oliva97} at
the pupil image position after the collimator lens, four images by
linearly polarized rays at 0$^{\circ}$, 90$^{\circ}$, 45$^{\circ}$
and 135$^{\circ}$ position angles (PAs), respectively, are recorded
on two 2k4k HPK CCDs simultaneously.
This enables us to obtain all three Stokes parameters for linear
polarization, i.e., $I,\ Q,\ U$, from only a single exposure.
Our observation started at 2009 Dec 8.41142 UT, $t=$149~s, 
which was automatically processed after receiving the
{\it Swift}/BAT Notice via GCN.
This is one of the earliest polarimetry to date, as far as we know.
We took ten 30~s exposures and then nine 60~s exposures through
a $15'\phi$ aperture mask and an $R$-band filter.
The observation finished at Dec 8.42458, $t=$1286~s.
Figure~\ref{fig:img091208b} shows a sample image of GRB~091208B
obtained with HOWPol.

\begin{figure}
  \begin{center}
    \begin{tabular}{c}
      \resizebox{85mm}{!}{\includegraphics{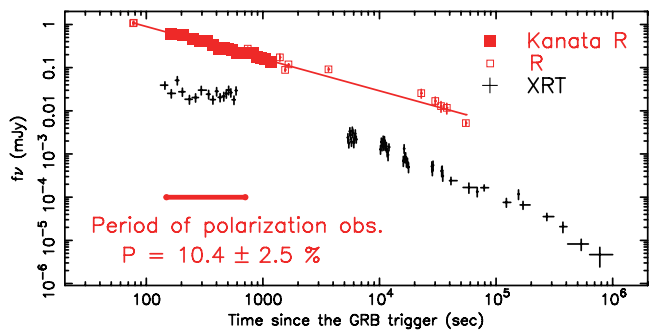}} \\
    \end{tabular}
    \caption{Optical and X-ray light curves of GRB~091208B.
Our optical and {\it Swift}/XRT data
are indicated by the filled squares and crosses, respectively.
Open squares are the optical data reported in GCN.
The solid lines are the best fitted power-law models for the optical light curve
(with the decay index of $\alpha_{\rm O} = -0.75 \pm 0.02$).
The thick horizontal bar at the left bottom part shows the period of our polarimetry. 
The derived polarization degree is also indicated. 
}
 \label{fig:lc091208b}
    \end{center}
\end{figure}

The raw data were reduced in a standard way for CCD aperture photometry.
For the photometric calibration, we used $R2$ magnitudes of
three nearby stars (C1--C3: USNO B 1068-0020023, 1068-0020019,
and 1069-0020340).
The optical light curve can be described with a single power-law form
(decay index $\alpha_{\rm O} = -0.75 \pm 0.02$), as shown in 
Figure~\ref{fig:lc091208b}.
For polarimetry, we could not use the sixth exposure (centered at
$t=$376~s) and all exposures after 13th ($t=$791~s)
because one polarization image of the GRB out of four falls into
the gap of the two CCDs due to slight telescope guiding error.
It reduces the number of the available frames for polarimetry to 11.
As for polarimetric calibration, we corrected for the instrumental
polarization of $P_{\rm instr}\simeq 3.9$~\%, predominantly caused by the
$45^{\circ}$-incidence reflection on the tertially mirror of the telescope.
The instrumental polarization has been modeled with an accuracy of 
$\Delta P_{\rm instr}\lesssim 0.5$ \% as a function of the hour angle and 
the declination (and also of the position taken in the field 
of view) of the object by systematic observations for unpolarized 
standard stars, and then checked by observations for strongly-polarized 
standard stars. 
In the case of GRB 091208B, it changed gradually with time from
$Q_{\rm instr}=-3.65$ \% to $-3.70$ \% and
$U_{\rm instr}=-1.05$ \% to $-0.88$ \% over the 11 exposures.
The detailed procedure and reliability of this `one-shot polarimetry'
will appear in a forthcoming paper (K. S. Kawabata et al., in preparation).

\begin{figure}
  \begin{center}
    \begin{tabular}{c}
      \resizebox{75mm}{!}{\includegraphics{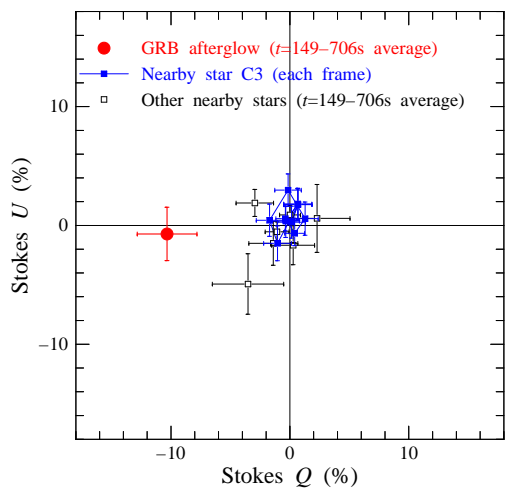}} \\
    \end{tabular}
    \caption{$QU$-diagram of the GRB afterglow and nearby stars.
For the bright comparison star C3, we demonstrate the frame-to-frame
variation of $Q$ and $U$, which suggests the residual systematic is 
negligible ($\lesssim 1$\%). For other stars we show time-averaged 
polarization at $t = 149 - 706\;$s.}
 \label{fig:nearbypol}
    \end{center}
\end{figure}

Since the S/N ratio of each single exposure is not sufficient for
polarimetry ($\Delta P\gtrsim 5$ \%), we combined the all 11 $Q$
and $U$ parameters to enhance the reliability.
We performed a traditional, statistic correction for the polarization 
bias in cases of low S/N as $P_{\rm real} = \sqrt{P^{2}-(\sigma_{P})^{2}}$
(\citealt{Serkowski58}, see also \citealt{Patat06}).
The derived polarization is $Q=-10.3\% \pm 2.5$\%\ and $U=-0.7\% \pm 2.2$ \%
(i.e., $P=10.4\% \pm 2.5$ \%\ and PA=$92^{\circ}\pm 6^{\circ}$).
The Galactic interstellar extinction indicates that the
interstellar polarization toward this GRB is negligibly small
($P_{\rm ISP}\leq 9 E_{B-V} = 0.5$ \% ; \citealt{Serkowski75, Schlegel98}).
To check the consistency, we obtained the polarization of nearby
stars ($\leq 4'$) brighter than or comparably faint to the GRB a
fterglow taken in the same frames, and plot them in $QU$-diagram
(Figure~\ref{fig:nearbypol}). 
Assumed that they are mostly Galactic normal stars having little 
or no intrinsic polarization, the diagram would support that 
the GRB afterglow has a significant polarization.

\section{X-ray Data: Reduction and Analysis}

We reduced the public data of XRT on {\it Swift} for GRB~091208B, using 
the XRT pipeline FTOOL {\it xrtpipeline} (Version:~2.3.3).  
The obtained 0.3--10 keV light curve is shown in Figure~\ref{fig:lc091208b}.
The first orbit data at $t=130$--$600$s exhibit possible bumps and mini-flares
(cf. \citealt{pagani10}).
The following orbit data can be described with a broken power-law form;
$F_{\nu} \propto t^{\alpha_{X,2}} (t \leq t_{\rm X,break})$ and ,
$F_{\nu} \propto t^{\alpha_{X,3}} (t > t_{\rm X,break})$.
We obtained the decay indices and the break time as listed in Table~\ref{tab:grb091208b_x}.

Based on the variability in the light curve,
we defined three phases for GRB~091208B, as described in Table~\ref{tab:grb091208b_x}.
For each phase, we obtained the time-averaged spectrum at 0.3--8.0 keV.
We fitted a power-law model having two absorption components to the X-ray data.
One component is the absorption in our galaxy and the other is the host galaxy.
For the former, the Galactic hydrogen column density has been derived as 
$N_{\rm H}^{gal} = 4.8 \times 10^{20}\; {\rm cm}^{-2}$ for the direction of 
GRB 091208B \citep{Schlegel98, Dickey90} and we fixed it. 
For the latter, we confirmed that the column density in the host galaxy 
$N_{\rm H}^{ext}$ is practically constant by our trial fitting and put
$N_{\rm H}^{ext}$ as a common parameter for all three phases. 
We show the result of our fitting in Table~\ref{tab:grb091208b_x}.
$N_{\rm H}^{ext}$ is derived as $(8.0\pm 2.1)\times 10^{21}$ cm$^{-2}$. 
It is noted that the spectral index $\beta_{\rm X}$ does not significantly 
change through the period of our optical polarimetry.

\begin{figure}
  \begin{center}
    \begin{tabular}{c}
      \rotatebox{270}{\resizebox{42mm}{!}{\includegraphics{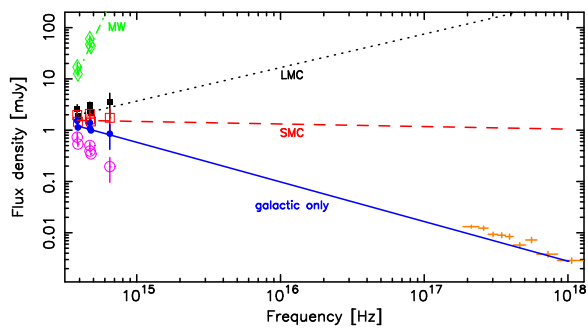}}} \\
    \end{tabular}
    \caption{SED of the Optical -- X-ray regime of GRB~091208B on $t=162-589$~s.
The crosses show the unabsorbed X-ray spectra. 
The purple open circles are raw optical data, and the blue small filled circles
are optical ones dereddened for the Galactic extinction only ($E_{B-V}=0.162$).
The red open squares, black filled squares and green open squares 
are optical data dereddened for the host extinction 
($N_{\rm H}^{ext}=8.0\times 10^{21}$ cm$^{-2}$) of the SMC, LMC and
MW models, respectively, in addition to the Galactic extinction.
The blue solid, red dashed, black dotted, and green dashed-dotted lines indicate 
the best-fitted power-law models of the individual dereddened optical data.
The blue solid line (with power-law index $\beta = -0.72 \pm 0.01$) appears
consistent with the X-ray data, too; however, this spectral model is not compatible
with the temporal evolutions of the optical and X-ray fluxes in the forward
shock synchrotron model (see \S~\ref{sec:model}).
}
 \label{fig:sed}
    \end{center}
\end{figure}

In Figure~\ref{fig:sed}, we show the spectral energy distribution (SED) 
from optical ($3.9 \times 10^{14}$~Hz) to X-ray band ($1.1 \times 10^{18}$~Hz)
at $t=162$--$589$~s.
The Galactic extinction of $E_{B-V}=0.162$ \citep{Schlegel98} has been
corrected for using a standard way. On the other hand, for the extinction 
within the host galaxy, we adopted three extinction curve models, 
the Milky Way (MW), the Large Magellanic Cloud (LMC), and the Small 
Magellanic Cloud (SMC) models, while the hydrogen column density 
($N_{\rm H}^{ext}=8.0\times 10^{21}$ cm$^{-2}$) is fixed 
\citep{Uehara10, Uehara11}.
Using one of these three models, we corrected the extinction in the 
optical bands. The extinction for the SMC model (or less than that) 
seems reasonable, since the temporal decay of the optical flux implies 
a soft spectrum in the synchrotron model. 
In this case, the flux density is 1.4 $\pm$ 0.3~mJy at 3.9$\times 10^{14}$~Hz.

\section{Discussion}

\subsection{Modeling of the Afterglow}
\label{sec:model}

The observed optical afterglow has a typical power-law light curve that is explained by 
the synchrotron emission from the forward shock propagating in the external medium
\citep{meszaros97,sar98grblc}. The decay index $\alpha_{\rm O} = -0.75 \pm 0.02$
is very likely to be the case $\nu_m < \nu_{\rm O} < \nu_c$ with the uniform-density 
external medium and $p \simeq 2.0$, where $\nu_{\rm O} \equiv 3.9 \times 10^{14}\;$Hz,
$\nu_m$ is the minimum injection frequency,  $\nu_c$ is the cooling frequency, and 
$p$ is the index of the electron energy distribution \citep[see Table 1 of][]{zhang04}. 
The X-ray emission in the second phase
also looks a typical power-law light curve, although we find that the above model 
for the optical emission cannot explain the X-ray emission simultaneously, as shown below.

The X-ray emission in the second phase shows 
$\alpha_{X,2} - 3\beta_{X,2}/2 = 0.48 \pm 0.24$. This can be interpreted as the case 
$\nu_X > {\rm max}(\nu_m, \nu_c)$ with $p \sim 2.1$. Thus one may consider the case
$\nu_m < \nu_{\rm O} < \nu_c < \nu_X$ in this phase. At some time earlier,
$\nu_c (\propto t^{-1/2})$ may have passed across $\nu_X$, which means that the
X-ray light curve can have a break and the decay index before the break is the same
as $\alpha_{\rm O}$. This enables us to have an X-ray light curve that is not brighter
than the observed X-ray emission in the first phase.
In this case, however, the optical and X-ray synchrotron emissions are in 
the same spectral segment, i.e., $\nu_m < \nu_{\rm O} < \nu_X < \nu_c$ with 
$\beta_{\rm O} = \beta_X \simeq -(p-1)/2 \sim -0.55$, 
which is not compatible with the observed joint spectrum in Figure~\ref{fig:sed}, 
in which the blue line has the spectral index $-0.72 \pm 0.01$.
The extrapolation of the optical flux at the higher frequencies with such a value of
$\beta_{\rm O}$ overwhelms the observed X-ray flux.

Considering that the observed X-rays at the first phase shows several flares,
it is reasonable that the optical emission is the forward shock
synchrotron radiation with $\nu_m < \nu_{\rm O} < \nu_c$, whereas the X-ray 
emission has different origins. Many models have been proposed for such anomalous
X-ray afterglows \citep[e.g.,][]{ghisellini07,kumar08,yamazaki09}.
In this case, the extrapolation of the optical flux at the higher frequencies has to have 
a cooling break at $\nu_{\rm O} < \nu_c(t \sim 300\;{\rm s}) \lesssim 10^{16}\;$Hz 
to suppress the emission below the observed X-rays. The other necessary conditions
for this model are the following: $\nu_m (t \sim 80\;{\rm s}) < \nu_{\rm O}$ and
$\nu_c (t \sim 5 \times 10^4\;{\rm s}) \gtrsim \nu_{\rm O}$ to have a single 
power-law light curve at the optical band; and $F_{\rm O} (t \sim 300\;{\rm s})
\simeq 1\;$mJy. The condition for $\nu_c$ can be summarized by using 
$\nu_c \propto t^{-1/2}$ as 
$5.0 \times 10^{15}\;{\rm Hz} \lesssim \nu_c (t\sim 300\;{\rm s}) \lesssim 10^{16}\;{\rm Hz}$.

The characteristic quantities in the forward shock synchrotron model with the uniform-density
external medium are given by \cite{granot02}
\begin{eqnarray}
\nu_m &\simeq& 7.1 \times 10^{10}\;E_{52}^{1/2} \epsilon_{e,-1}^2 f_{-1}^2 \epsilon_{B,-2}^{1/2}
t_{\rm days}^{-3/2}\;{\rm Hz},\\
\nu_c &\simeq& 6.7 \times 10^{15}\;E_{52}^{-1/2} \epsilon_{B,-2}^{-3/2} n_0^{-1} 
t_{\rm days}^{-1/2}\;{\rm Hz}, \\
F_{\nu_m} &\simeq& 0.91\;E_{52} \epsilon_{B,-2}^{1/2} n_0^{1/2}\;{\rm mJy},
\end{eqnarray}
where $E_{\rm iso} = 10^{52} E_{52}\;$erg is the total isotropic energy of the blast wave, which
may be comparable to $E_{\gamma,{\rm iso}} \simeq 1 \times 10^{52}\;$erg (see Section~\ref{sec:intro}), 
$n = 1 n_0\;{\rm cm}^{-3}$ is the density of the
external medium, and $\epsilon_e = 10^{-1} \epsilon_{e,-1}$ and $\epsilon_B = 10^{-2} \epsilon_{B,-2}$
are the fractions of the shocked energy carried by the electrons and the magnetic field, respectively.
The case of $p \simeq 2.0$ involves the factor $f = 1/\ln (\gamma_M/\gamma_m)$, where 
$\gamma_M$ and $\gamma_m$ are the maximum and minimum injection Lorentz factors of 
electrons, respectively, and $f$ typically has a value of $\sim 10^{-1}$. The optical emission 
is calculated as $F_{\rm O} = F_{\nu_m} (\nu_{\rm O}/\nu_m)^{-(p-1)/2}$. The necessary 
conditions given above are translated into
\begin{eqnarray}
E_{52}^{1/2} \epsilon_{e,-1}^2 f_{-1}^2 \epsilon_{B,-2}^{1/2} < 0.15, \\
E_{52}^{1/2} \epsilon_{B,-2}^{3/2} n_0 \simeq 11.4\; y, \\
E_{52}^{5/4} \epsilon_{B,-2}^{3/4} \epsilon_{e,-1} f_{-1} n_0^{1/2} \simeq 1.2,
\end{eqnarray}
where we define a numerical factor $1 \lesssim y \lesssim 2$. The equations are reduced into
$\epsilon_{e,-1} f_{-1} \simeq 0.35\; E_{52}^{-1} y^{-1/2}$, 
$\epsilon_{B,-2} < 1.6\; E_{52}^3 y^2$, and
$n_0 > 5.5\;E_{52}^{-5} y^{-3}$.
These equations indicate that the optical emission can be understood as the forward shock
synchrotron emission with the typical parameter values \citep{panaitescu02}.

\subsection{Implications from Optical Polarization}

We have measured the optical polarization in the early forward shock emission of this burst
as $P = 10.4\% \pm 2.5\%$. 
Since this is a mean value over a long observation time $t=149 - 706\;$s, 
the actual polarization degree might have been either unusually constant or temporally 
much higher.
This result disfavors the afterglow model in which the magnetic field 
coherent scales are plasma skin depths in the emission region, as explained below.

The magnetic field can be amplified by the plasma instabilities at the shock
front, such as Weibel instability. This can produce the strong fields with random directions on the 
skin depth scales, which are very tiny compared with the size of the observable region of the shock
\citep{medvedev99}. The field directions are possibly constrained to be parallel to the shock
plane. 
In this case, however, the observed polarization reaches a maximum value around the 
jet break time (typically around $t \sim 1\;$day) in general 
\citep{sari99,ghisellini99,rossi04}. 
Our measurement of the early optical polarization which is higher than the typical
late-time polarization, $\sim 1$--$3$\%\ (at $t \sim 1\;$day; \citealt{covino04}) 
is thus not consistent
with this mechanism.\footnote{The jet with non-uniform angular energy distribution 
and with the plasma-scale fields can produce high degree of polarization 
at any times \citep{nakar04}, although it may not produce such a smooth light curve 
as we observed. The residuals of our optical data at $t=149-706\;$s from the
power-law model are all within the $1 \sigma$ level.} 
Some polarization angle data at late-time are also incompatible with this mechanism
\citep{lazzati04,lazzati06}. 
Furthermore, the numerical simulations of 
collisionless shocks suggest that the fields produced by the plasma instabilities decay fast 
and do not survive on the large scales corresponding to the emission region 
\citep{sironi09,gruzinov99}.

We suggest that magnetohydrodynamic instabilities may be viable for the 
magnetic field origin in the emission region, rather than the plasma instabilities.
If the shock sweeps inhomogeneous external medium, multiple vorticities arise downstream 
of the shock due to the growth of the Richtmyer-Meshkov instability, increasing the 
field strength \citep{sironi07, inoue11}.
The numerical simulations show that this mechanism can produce strong fields random on large scales
that are roughly comparable to the density fluctuation scale of the external medium.
Such large-scale fields decay rather slowly, which may survive in the entire emission region.
In this case the observed polarization scales as $P \sim 70\%/\sqrt{N}$,
where $N$ is the number of the coherent field patches in the observable region 
$\theta \lesssim 1/\Gamma$ \citep{gruzinov99}. Our measured polarization degree means
$N \sim 50(P/10\%)^{-2}$. 
Since the radius $R$ and the Lorentz factor $\Gamma$ of the forward shock are estimated to be
$R \simeq 4.6 \times 10^{17}\;E_{52}^{1/4} n_0^{-1/4} t_{\rm days}^{1/4}\;$cm, and 
$\Gamma \simeq 4.7\;E_{52}^{1/8} n_0^{-1/4} t_{\rm days}^{-3/8}$ \citep{granot02},
we may estimate the typical scale of the field coherent patch at $t \sim 300\;$s as 
$l_p \sim (R/\Gamma)/\sqrt{N} \sim 4 \times 10^{14}\;E_{52}^{1/8} n_0^{-1/8} (P/10\%)\;$cm.
The coherent patches with such scales might be formed by the density fluctuations of 
molecular cloud cores or H\textsc{i} clouds in the interstellar medium, of which 
the sizes are typically $\sim 10^{16} - 10^{17}\;$cm.

In this model, the polarization angle randomly changes with time, and the
polarization degree scales as
\begin{equation}
P \propto 1/\sqrt{N} \propto l_p (R/\Gamma)^{-1} \propto l_p E^{-1/8} n^{1/8} t^{-5/8}.
\end{equation}
At $t \sim 1\;$day, this burst may show $P \sim 0.3\%$, if $l_p$ is roughly constant
through the shock propagation. 
This value is comparable to (or somewhat smaller than) the typical 
late-time polarization, $\sim 1-3\%$.
If the shock propagates in the magnetized wind from the progenitor star, the polarization 
from the ordered field, which has roughly constant degree and angle, contributes to 
the net polarization \citep{granot03}, although the light curve of this burst implies that the 
density profile of the external medium is constant ($n \propto r^0$) rather than the wind
type ($n \propto r^{-2}$) (see Section~\ref{sec:model}). 

The result of \citet{Mundell07}, a $2\sigma$ upper-limit $P < 8$\% for GRB~060418, 
might imply that $l_p$ was several times smaller than that for the GRB 091208B case.
Anyway, we need more data of polarization and also their time variation for other 
bursts, which would further constrain the field structure in the emission region 
and the nature of the environment of GRBs.

\begin{acknowledgements}
This work is supported by JSPS Research Fellowships for Young
Scientists (TU, KT, RI, MS, and MY) and by the Grant-in-Aid 
for Scientific Research from JSPS (17684004, 19047003, 23244030, 23340048, and 60372702).
\end{acknowledgements}

\begin{table*}
 \caption{Decay and spectral indices of the X-ray afterglow of GRB~091208B}
\label{tab:grb091208b_x}
\centering
\begin{tabular}{cccc}
\hline\hline
Phase       & $T$ (s)                            & $\alpha_{\rm X} \dag$ & $\beta_{\rm X} \ddag $  \\ \hline
\it{First}  & 130--600                           & $-0.18\pm 0.15$       & $-0.97\pm 0.14$         \\
\it{Second} & $5300$--$3.1\times 10^5$           & $-1.10 \pm 0.12$      & $-1.05^{+0.14}_{-0.13}$ \\
\it{Third}  & $3.1\times 10^5$--$1.0\times 10^6$ & $-2.3^{+1.6}_{-1.2}$  & $-0.82_{-0.42}^{+0.30}$ \\
\hline
   \multicolumn{4}{@{}l@{}}{\hbox to 0pt{\parbox{100mm}{\footnotesize
        The uncertainties show the 90\% confidence levels of the parameters.  \\
       $\dag$ X-ray decay index. \\
       $\ddag$ X-ray spectral index($\chi^2$/d.o.f = 60.9/78). \\
     }}}
\end{tabular}
\end{table*}

\end{document}